\documentclass[reprint,showpacs,twocolumn,superscriptaddress,aps]{revtex4-2}

\usepackage[utf8]{inputenc}
\usepackage[american,british]{babel}
\usepackage[T1]{fontenc}
\usepackage[pdftex]{graphicx}  
\usepackage{graphicx}
\usepackage{xcolor}
\usepackage{mathtools}
\usepackage{bbm}
\usepackage{subfigure}
\usepackage{dcolumn}
\usepackage{braket}
\usepackage{bm}
\usepackage{amsmath,amsthm,amssymb}
\usepackage{color}
\usepackage{verbatim}
\usepackage{comment}

\definecolor{Red}{HTML}{E53E30}  
\definecolor{Green}{HTML}{00AD69}  
\definecolor{Blue}{HTML}{2171b5}
\definecolor{Purple}{HTML}{652F6C}  

\usepackage[colorlinks,citecolor=blue,linkcolor=blue,urlcolor=blue,hyperindex]{hyperref}

\newcommand{\mdm}[1]{#1}

\newcommand{\vv}[1]{#1}
\newcommand{\vvv}[1]{#1}

\usepackage{dsfont}
\usepackage{physics}

\begin{document}
\title{Topological Kolmogorov complexity and the Berezinskii-Kosterlitz-Thouless mechanism}

\author{Vittorio Vitale}
\email{vittorio.vitale@lpmmc.cnrs.fr}
\affiliation{International Centre for Theoretical Physics (ICTP), Strada Costiera 11, 34151 Trieste, Italy}
\affiliation{SISSA, via Bonomea 265, 34136 Trieste, Italy}
\affiliation{Université Grenoble Alpes, CNRS, Laboratoire de Physique et Modélisation des Milieux Condensés (LPMMC), Grenoble 38000, France}

\author{Tiago Mendes-Santos}
\affiliation{Theoretical Physics III, Center for Electronic Correlations and Magnetism,
Institute of Physics, University of Augsburg, 86135 Augsburg, Germany}

\author{Alex Rodriguez} 
\affiliation{International Centre for Theoretical Physics (ICTP), Strada Costiera 11, 34151 Trieste, Italy}
\affiliation{Dipartimento di Matematica e Geoscienze, Universit\'a degli Studi di Trieste, via Alfonso Valerio 12/1, 34127, Trieste, Italy}

\author{Marcello Dalmonte}
\affiliation{International Centre for Theoretical Physics (ICTP), Strada Costiera 11, 34151 Trieste, Italy}
\affiliation{SISSA, via Bonomea 265, 34136 Trieste, Italy}

\date{\today}

\begin{abstract}
Topology plays a fundamental role in our understanding of many-body physics, from vortices and solitons in classical field theory, to phases and excitations in quantum matter. Topological phenomena are intimately connected to the distribution of information content - that, differently from ordinary matter, is now governed by non-local degrees of freedom. However, a precise characterization of how topological effects govern the complexity of a many-body state - i.e., its partition function - is presently unclear. In this work, we show how topology and complexity are directly intertwined concepts in the context of classical statistical mechanics. In concrete, we present a theory that shows how the \emph{Kolmogorov complexity} of a classical partition function sampling carries unique, distinctive features depending on the presence of topological excitations in the system. We confront two-dimensional \vv{Ising, Heisenberg} and XY models on several topologies, and study the corresponding samplings as high-dimensional manifolds in configuration space, quantifying their complexity via the intrinsic dimension. While for the Ising \vv{and Heisenberg} model the intrisic dimension is independent of the real-space topology, for the XY model it depends crucially on temperature: across the Berezkinskii-Kosterlitz-Thouless (BKT) transition, complexity becomes topology dependent. In the BKT phase, it displays a characteristic dependence on the homology of the real-space manifold, and, for $g$-torii, it follows a scaling that is solely genus dependent. We argue that this behavior is intimately connected to the emergence of an order parameter in data space, the conditional connectivity, that displays scaling behavior. Our approach paves the way for an understanding of topological phenomena emergent from many-body interactions from the persepctive of Kolmogorov complexity.
\end{abstract}

\maketitle

\begin{figure*}
    \centering
    \includegraphics[width=0.9\linewidth]{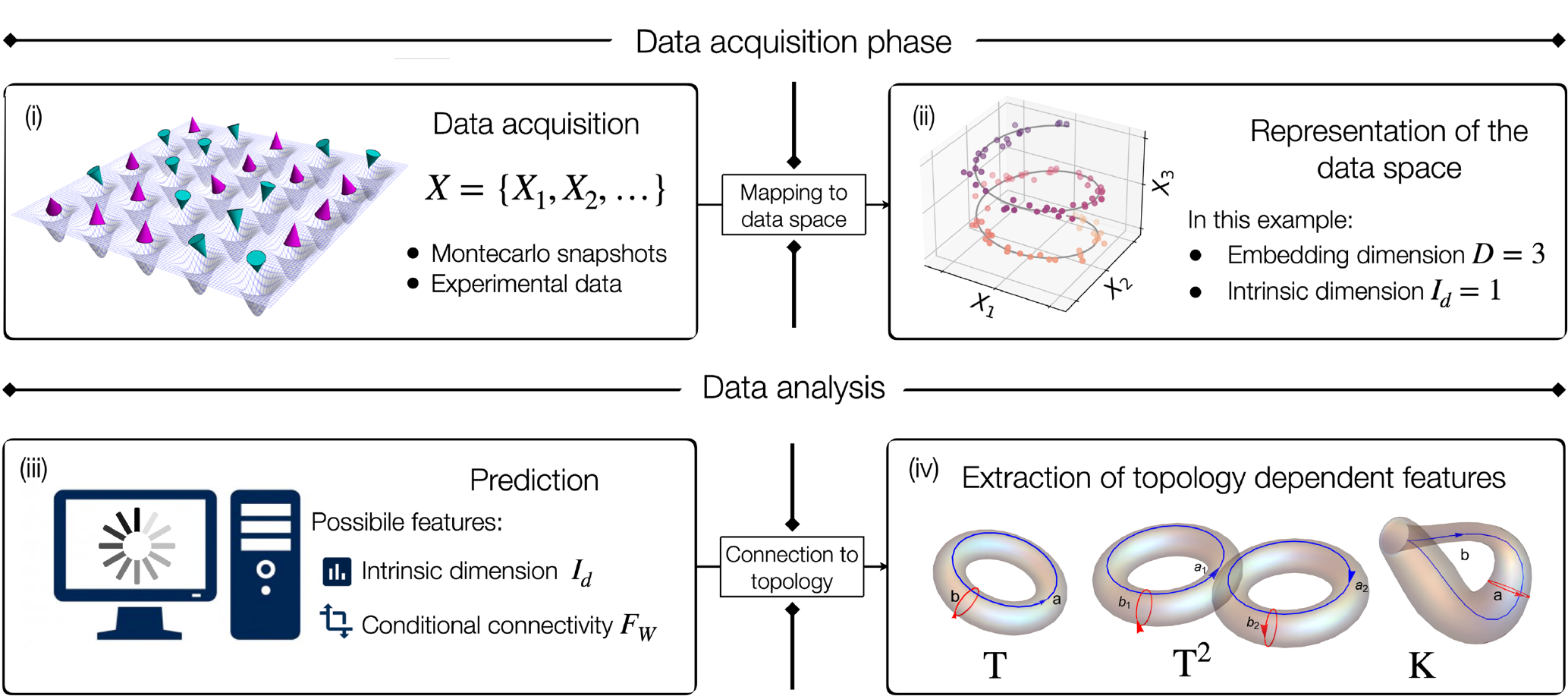}
    \caption{{\bf Guideline to connect complexity of classical partition functions to topological phenomena.} (i) We collect snapshots of the physical system - either Monte Carlo configurations or data from experiments. (ii) We map the configurations \vv{to} a $D$ dimensional (embedding dimension) data space. (iii) We extract the features of the data set, e.g.intrinsic dimension ($I_d$), conditional connectivity ($F_{\mathbf{w}}$). (iv) We connect the findings to the manifolds the system lives in. Here we represent a torus (T), a 2-torus (T$^2$) and a Klein bottle (K).}
    \label{fig:summary} 
\end{figure*}

Complexity and topology are two cornerstones in our interpretation and understanding of phenomena and mechanisms in diverse fields of science \cite{ladyman2013complex}. One particular domain where the two are at play is many-body physics: there, one is interested in how collective behavior emerges from the interplay of a large number of microscopic components, thus enabling effective descriptions that only use a finite number of variables. At the information-theoretic level, this implies a huge reduction of the Kolmogorov complexity corresponding to a given description of the physical system: a handful of parameters are sufficient to capture essential features of the corresponding physical phenomenon \cite{li2008introduction}. The prototypical example of such reduction of complexity is critical behavior, where correlation functions are precisely determined by a few parameters: the critical exponents~\cite{seancarrol}.

In recent years, thanks to dramatic advances in the field of non-parametric unsupervised learning methods~\cite{mcinnes2018umap,kaya2019deep,kulis2013metric}, it has become possible to make quantitative statements about the complexity of many-body processes in the context of molecular science and, broadly speaking, quantum chemistry and physics \cite{Carleo2019,Carrasquilla2020,Mehta20191}. In those domains, estimating complexity is particularly important to understand the interplay between the multitude of degrees of freedom, and, eventually, to help identify those that are most relevant for a physical phenomenon to occur. This is intimately related to `feature selection'~\cite{jovic2015review,cai2018feature,deng2019feature,glielmo2022ranking} or `dimensional
reduction'~\cite{van2008visualizing,bengio2013representation}, i.e. discarding features that appear irrelevant or redundant, or finding a representation of the data with few variables seen as complicated functions of the original ones. The same tools have been recently applied to critical phenomena \cite{Wang2016,Wenjian2017,Wetzel2017}, where they have been shown to lead to qualitative insights \vv{on} how complexity is tied to critical behavior~\cite{TiagoPRX}. Instead, very little is known about how complexity is dictated by topological phenomena: given a partition function, is it possible to infer that non-local degrees of freedom are at play by solely looking at its complexity? Beyond theoretical interest, addressing such a question can also be of experimental relevance to a variety of systems, including cold atom gases~\cite{hadzibabic2006berezinskii}, arrays of Josephson junctions~\cite{han2014collapse}, and quantum annealing architectures~\cite{king2018observation}, where topological phenomena have been recently demonstrated, and where single-site resolution is often possible.

Here, we show that, in classical statistical mechanics, the topological origin of the Berezinskii-Kosterlitz-Thouless~\cite{drouin2022kosterlitz} mechanism is tied to the emergence of a topological contribution to the Kolmogorov complexity characterizing the partition function. This relation is uniquely determined by the nature of the manifold upon which the statistical mechanics problem is defined, and is in sharp contrast to what happens for models governed by local order parameters - whose complexity is unrelated to the manifold topology. 

First, we provide a theoretical framework that argues how complexity decreases as the dimension of the homology group of the real-space manifold increases: this reflects the fact that, the more complex the real-space manifold, the more constrained low-energy excitations of the system are. We then analyze the partition function of \vv{XY, Ising and Heisenberg models} on an array of two-dimensional manifolds - orientable as well as non-orientable. For each manifold, we estimate the Kolmogorov complexity utilizing the intrinsic dimension, a measure of the minimal number of degrees of freedom required to describe a data structure. We then define as topological Kolmogorov complexity of a manifold the difference between the complexity of the partition function on that manifold, and the one on the 1-torus \vv{($T^1$)}. 

In the BKT phase of the XY model, we observe \vv{that the complexity is strongly dependent} on the manifold the model is defined on. In agreement with our theory, such complexity is dictated by the first homology group and satisfies a phenomenological finite-size scaling. Our results indicate that, while information is non-locally encoded in real space (i.e. in winding numbers corresponding to vortex excitations), it is remarkably {\it local} in data space: this is shown by the emergence of a local order parameter \vv{-- in data space --}, that we dub \emph{conditional connectivity}, and that displays scaling behavior across the BKT transition. The identification of such order parameters provides a key, intuitive connection between homology and complexity. In contrast, for the case of the \vv{Ising and Heisenberg model}, we find no relation between complexity and real-space topology, again in agreement with our theory. 

Our findings demonstrate a first, direct link between topology and complexity of a many-body description, that parallels recent efforts in both classical~\cite{tu2017universal,tang2017universal} and quantum~\cite{kitaev2006topological,levin2006detecting,Fromholz2020,Micallo2020} statistical mechanics in terms of entropy content. Based on the impact of complexity in other disciplines, this new connection \vv{opens the door} towards a transfer of concepts and methods between different fields.

\section{Kolmogorov complexity of partition function samples}\label{sec:Theory}

Kolmogorov complexity is the fundamental measure of algorithmic complexity and has found widespread application in diverse fields of science (for an overview, see Ref.~\cite{li2008introduction}). 
It corresponds to the length of the shortest computer program that produces a given output, that, in our case, corresponds to the data set built with the configurations sampled along the MC simulations.
From an alternative viewpoint, it characterizes the capability of the output (the data set) to be compressed. In the following, our goal is to connect this information-theoretic viewpoint to topological phenomena. To establish the relation between topological features and complexity, we formulate a four-step procedure (summarized in Fig.~\ref{fig:summary}), that encompasses {\it (i)} data acquisition, {\it (ii)} proper representation and {\it (iii)} data mining analysis in configuration space, {\it (iv)} repeated over different manifolds to single-out topological features.

{\it (i) Data structures of partition functions. -} The first element of our discussion concerns the identification of the data set we target to describe a physical phenomenon. For an equilibrium classical statistical mechanics model, the starting point is very natural: the partition function. While the latter cannot be measured experimentally (nor, in many cases, directly computed), we focus on its sampling. In particular, labeling as $X_i$ an element of the configuration space, our target data set is a collection of $N_r$ such elements:
\begin{equation}\label{eq:datasets}
    X=\left\{ X_1,X_2,\dots ,X_{N_r} \right\},
\end{equation}
with $N_r\ll D_C$, where $D_C$ is the total dimension of the configuration space (e.g., $2^N$ for a collection of $N$ Ising spins). We will discuss the case $X_i\neq X_j$, which, as we argue below, is typical of large many-body systems. The data space $X$ contains information about arbitrary rank correlation functions, so it is expected to provide a full characterization of physical properties. Importantly, it is immediately available both in numerical (e.g., Monte Carlo) and real experiments. 

{\it (ii) Intrinsic dimension and complexity. - } 
Once the data space is defined, it is important to find a suitable measure of information compression that serves as an estimator for the Kolmogorov complexity, $K(X)$. Commonly, data points in a data set are represented as points in a space whose dimension is the number of features needed to describe each sample, which is called \emph{embedding dimension}. \vv{The correlation between} the data points determines the manifold in which the data lie, inducing a structure whose dimension is typically much smaller than the embedding one. This is known as \emph{intrinsic dimension} and denoted by $I_d$. For data sets whose local density is approximately constant at the scale of nearest neighbors, the $I_d$ corresponds roughly to the minimum number of variables needed to describe a data set, and is thus a direct estimator of Kolmogorov complexity. This connection has been shown in graph theory using one of the possible estimators of the $I_d$: the Hausdorff dimension \cite{mendes2023wave,staiger1993kolmogorov}.

The relation between intrinsic dimension and complexity has found applications in various fields. In molecular science, $I_d$ is used to analyze the complexity of large data sets of molecular structures, such as protein structures, to gain insights into the underlying structural motifs that govern the molecule functions~\cite{butler2018machine,facco2019intrinsic}. In computer vision, it is used to analyze images and videos, identifying the most important features and patterns relevant for tasks such as image classification or object detection~\cite{gong2019intrinsic,pope2021intrinsic}. In physics, $I_d$ is used to characterize both quantum and classical systems~\cite{Verdel2023BECS,TiagoPRX,TiagoPRXQ,Xhekdata}, estimating the number of degrees of freedom required to describe a given physical system across phase transitions, and its universality has been numerically proved.

{\it (iii) Estimators for the intrinsic dimension. -} Different approaches have been proposed to estimate $I_d$ ~\cite{camastra2016intrinsic,campadelli2015intrinsic,van2008visualizing,kramer1991nonlinear,wold1987principal}. The technique used here, the TWO-NN method~\cite{facco2017estimating}, relies on the statistics of distances between nearest-neighbors elements in the data set. 
The assumption of such approaches is that nearest-neighbor points can
be considered as uniformly drawn from $I_d$-dimensional hyperspheres. Hence, it is \vv{possible} to set
relations between the $I_d$ and the statistics of neighboring
distances and is particularly suitable for
non-linear manifolds. In the TWO-NN, one calculates the ratio $\mu_i=r_i^{(2)}/r_i^{(1)}$ for each point $X_i$ in the data set, where $r_i^{(1)}$ and $r_i^{(2)}$ are the nearest-neighbor and the next-nearest-neighbor distances, respectively. Under the assumption above, it can be proved that $\mu_i$ is distributed according to 
\begin{equation}
    f(\mu)=I_d \mu^{-I_d-1}.
\end{equation}
Therefore, $I_d$ can be estimated from the cumulative distribution function $P(\mu)$ as
\begin{equation}
    I_d=-\frac{\ln{\left[1-P(\mu)\right]}}{\ln{\mu}}.
\end{equation}
\vv{An example is shown in Fig.~\ref{fig:Paretocheck} for three cases: the estimation of $I_d$ corresponds to the slope of the straight lines.}
It is worth mentioning that $I_d$ is a scale-dependent quantity. In particular changing the value of $N_r$, number of configurations in the data set, amounts to changing the value of $I_d$ but not its functional dependence. Also, as mentioned above, $I_d$ has been utilized to investigate classical and quantum phase transitions because it shows a singular behavior in the vicinity of phase transition due to the change of correlation between the points in the data set. In essence, certain phases (e.g. the ferromagnetic phase in the Ising model or the BKT phase in the XY model \cite{TiagoPRX}) are characterized by configurations that are strongly correlated as they display the same physical properties. Conversely, outside these ordered phases, the data become lesser correlated the further we move away from the critical point (from a network theory viewpoint, they can be drawn from an Erd\"os-Renyi network). These effects reflect on the behavior of $I_d$ in parameter space.

Finally, it is worth commenting on the definition of distance to be used in the TWO-NN method. The choice of a proper metric is guided by the fact that it is non-negative, equal to zero only for identical configurations, symmetric, and satisfying the triangular inequality. In this work, we will use the Euclidean metric to estimate the distance between points, as a most natural one to impose a geometry in data space. Other suitable distances may be utilized equivalently in the present case.

{\it (iv) Topological Kolmogorov complexity}. We now need a criterion to determine whether complexity and topology `talk to each other' - that is, whether the complexity of a partition function can reveal the fact that information is carried by topological excitations, or not. Inspired by similar reasoning at the basis of topological field theory, we use the following approach: given a model, we input in two distinct (in a way we specify below) topologies and ask ourselves whether and how the complexity of partition function sampling has changed.

For simplicity, we consider a BKT phase (our arguments are immediately extended to cases with different topological defects, such as, e.g., liquid crystals~\cite{nakahara2018geometry}). 
Our starting point is the partition function of an XY model on a manifold with 1-torus topology, whose sampling and Kolmogorov complexity we denote as $X^0$ and $K(X^0)$, respectively. We then compare the latter with the one obtained from the sampling $X$ of \vv{the} partition function of the same model but on a manifold with different topology. 

The rationale behind this is the following~\footnote{We note, en passant, that this reasoning is loosely inspired by the one used in Ref.~\cite{levin2006detecting} to identify a topological contribution to entropy in quantum mechanical systems, with the key difference that here we are not referring to partitions, but rather, different manifolds.}. When topological degrees of freedom carry the information \vv{in} the system, i.e. at low temperatures in the XY model, a new set of constraints appears such as the winding numbers ($\bf{w}$) of these topological excitations. The paths on which independent winding numbers can be calculated, which also represent different and independent constraints on the partition function, are strictly related to the particular manifold taken into account. These paths are shown in Fig.~\ref{fig:summary}iv), where different windings are represented on 3 manifolds: a 1-torus \vv{($T^1$)}, a 2-torus \vv{($T^2$)} and a Klein bottle. The presence of different constraints must affect the complexity of the sampling of the partition function. Therefore, this effect is expected to be visible by studying its $I_d$. 

In order to identify topology-related information content, we then posit the 1-torus as a reference manifold, and we identify as $K_{\text{topo}}^{(\mathcal{M})}$ the \emph{topological Kolmogorov complexity} of the problem defined on a manifold $\mathcal{M}$ as the difference of the intrinsic dimension computed from the partition function obtained on $\mathcal{M}$, and that on the 1-torus:
\begin{equation}
    K_{\text{topo}}^{(\mathcal{M})} = K^{(\mathcal{M})} - K^{(T)}.
\end{equation}
In the following, we show numerically that this quantity strikingly \vv{distinguishes} phenomena which \vv{do} have a topological origin, from those that \vv{do} not. Most importantly, the former displays a universal scaling behavior for $g$-torii and a characteristic connection to the first homology groups for more complicated closed manifolds, demonstrating an unambiguous connection between topology and complexity.

\section{Models}

In order to illustrate our ideas above, we utilize \vv{three} models of classical statistical mechanics: the Ising, \vv{the Heisenberg model}, and XY model, as anticipated before. 
The 2D XY model displays a (BKT) topological phase transition associated with the emergence of topological defects as vortices, at temperature $T_{BKT} \simeq 0.8933$~\cite{berezinsky32destruction,kosterlitz1973ordering,sandvik2010computational}. The physical degrees of freedom are represented by rotors on the plane and the energy of the system can be written as follows
\begin{equation}
    E(\vec{\theta})=-\sum_{\langle i,j \rangle }\vec{S}_i\vec{S}_j,
\end{equation}
where $\vec{S}_i=(\cos{\theta_i},\sin{\theta_i})$ and $\theta \in [0,2 \pi[$. The configurations are represented by
\begin{equation}
    \vec{\theta}=\left\{ \cos{\theta_1},\sin{\theta_1},\dots,\cos{\theta_N},\sin{\theta_N}\right\},
\end{equation}
where $N$ is the total size of the system.\\

\vv{In order to compare the topological character of excitations of the XY model with a case where such phenomenology is absent, we investigated the classical Ising model and the classical Heisenberg model in two dimensions.}

The Ising model is described by
\begin{equation}
    E(\vec{s})=-\sum_{\langle i,j \rangle }s_i s_j
\end{equation}
where the degrees of freedom are spins $s_i =\pm 1$ and the sum runs on the nearest-neighboring bonds of a lattice. The
configuration states are defined as
\begin{equation}
    \vec{s}=\left\{ s_1,\dots,s_N\right\},
\end{equation}
where N is the number of spins in the lattice.
This model exhibits a second-order phase transition
characterized by the breaking of $\mathbb{Z}_2$ symmetry at the critical
temperature $T_c=2/ \ln(1+\sqrt{2})$. 

\vv{The Heisenberg model is described by
\begin{equation}
    E(\vec{s})=-\sum_{\langle i,j \rangle }\vec{S}_i \vec{S}_j
\end{equation}
where the sum runs on the nearest-neighbors and $S_i=\left(\sin{\theta_i}\cos{\phi_i},\sin{\theta_i}\sin{\phi_i},\cos{\theta_i}\right)$ is a three-components spin, with $\theta \in [0, \pi]$ and $\phi \in [0, 2\pi]$.  This statistical mechanics model displays no phase transition in temperature.}

\vv{We utilize the Ising model and the Heisenberg model as a benchmark: since they do not display a topological phase transition, they can be exploited to verify that the behavior of the Kolmogorov complexity in the 2D XY model is due to the particular topological features of the latter, and not by some other, unwanted (e.g. finite volume) effect.}

\section{Complexity and homology}\label{sec:complexityandhomology}

\subsection{Theory: winding numbers and information compression}\label{sec:windingnumberandinformation}

Before embarking on a quantitative determination of the intrinsic dimension in spin models, it is instructive to present a heuristic argument that anticipates the connection between topology and complexity.

Let us consider the data set $X$ introduced in Eq.~\eqref{eq:datasets}. The complexity of the data set $X$, i.e. $K(X)$, points out what is the information stored in the data, hence how much information content can be encoded by the system. When constraints on the manifold are enforced, we expect it to be reflected in a decrease of complexity as one can attach to the snapshots non-trivial labels, that we denote with $W_j$, $j=1,\dots,N_r$. In formulas, $K(X)$ can be written as the difference between the complexity of the bare data set $X^{0}$ and the contribution due to the presence of labels $W_j$ associated with each $X_j$:
\begin{equation}
    K(X)=K(X^{0})-K(W|X^{0}).
\end{equation}
This observation is intimately related to the topology of the manifold the data set lives in. Let us first illustrate this consideration with a simple example. We consider a set of spin variables describing a strongly ordered Ising-ferromagnetic state on a square lattice defined on a 1-torus. The configurations of the spins will be everywhere analogous to the one obtained by a lattice with open boundary conditions (OBC), with the exception of the constraints at the boundaries of the plane. This implies that the additional condition of periodic boundary conditions will reduce the complexity of the data set (physically, this is a geometrical effect, related to the fact that the open boundary conditions make magnetic order weaker in a finite region of space~\cite{henkel1999conformal}). However, when closed manifolds are considered, we do not expect any difference in the thermodynamic limit: the samples obtained from an Ising-ferromagnet will have the same complexity regardless of the underlying manifold.

\begin{figure*}
    \centering
    \includegraphics[width=\linewidth]{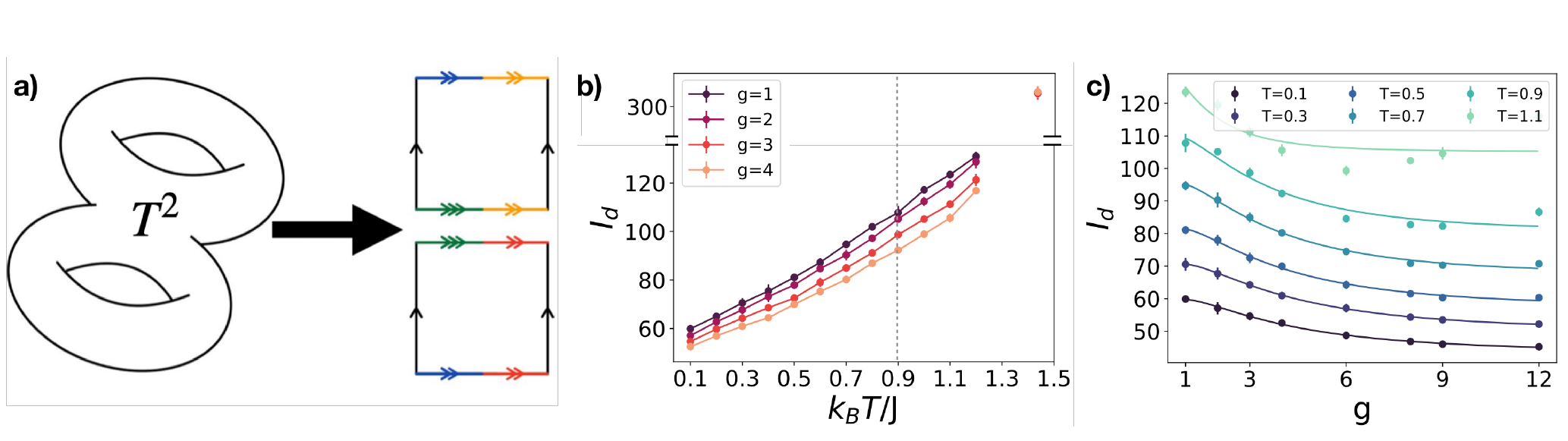}
    \caption{{\bf Plot of $I_d$ for $g$-torii}. We consider the XY model on a square lattice. a) Diagram of the construction of the 2-torus (T$^2$),  edge with same color and arrows must be \vv{connected} with periodic boundary conditions. b) $I_d$ of the two-dimensional XY model as a function of temperature $T$, for different increasing value of the genus $g$ (top to bottom); c) scaling of $I_d$ as a function of the genus $g$, the lines correspond to Eq.~\eqref{eq:genus_scaling}, for increasing temperature going from bottom to top. In both panels we consider a square lattice with linear dimension $L=72$. We average our results over 10 simulations with $N_r=5000$ different configuations. \vv{The error bars are estimated as the standard deviation of the mean.} \vvv{ The dashed vertical line marks the critical point.}}
    \label{fig:fig2}
\end{figure*}

The situation is expected to be drastically different if the magnetic state has a BKT origin. There, low-energy excitations are vortices: those cannot be distinguished by local operators, and are instead labelled by winding numbers. This implies that, at low temperatures, the space of configurations that is spanned by the partition function is strongly constrained by the fact that vortices must have specific \vv{windings}. This is in strong contrast with \vv{what} happens at higher temperatures, where vortices are unbound and free to propagate (and the concept of winding ultimately \vv{loses} its significance as configurations with very strong local spin fluctuations start being present).

\mdm{\paragraph*{Low-temperature expansion. -} 
In the low temperature phase, it is possible to make concrete, quantitative predictions on the above mentioned picture. Let us consider the data set obtained by sampling the XY partition function for $T\ll T_c$, with $N_r\gg1$, and on a closed manifold with homology group $\mathcal{G} = \bigoplus_{j=1}^{k} \mathbb{Z}_{\alpha_j} $. In this regime, a large amount of configurations will have all spins aligned in a given direction -- all of them in the extreme case $T=0$. Such cluster has intrinsic dimension $\sim 1$, and the average distance between states is $\propto \sqrt{N}/N_r$~\footnote{Let us remind the reader that the euclidean distance between two configurations $\vec{\theta^i}, \vec{\theta^j}$ reads~\cite{TiagoPRX}:
\begin{equation}
    d(\vec{\theta^i}, \vec{\theta^j}) = \sqrt{2 \sum_{k=1}^{N} (1-\vec{S_k^i}\vec{S_k^j})}.
\end{equation}}.
This cluster, that we refer to as $T=0$ cluster, has constant curvature: in fact, it can be (locally) thought of as a straight line in data space, due to the fact that all spins are described by a single angle $\theta$.}

\mdm{Importantly, these are not the only configurations that are sampled. A finite fraction of them, representing low-energy excitations, is also present. They can be divided into two classes: $(i)$ configurations that host local spin excitations, at energy difference $\delta\epsilon$ of order 1 (and, thus, vanishing energy density above the lowest energy states), and $(ii)$ topological excitations, also at finite energy difference above the lowest energy state, and with a defined set of winding numbers. Note that the fact that topological excitations, despite having wildly different spin configurations with respect to the fixed direction case, still have vanishing energy density is particularly relevant here. }

\mdm{The overall data structure, and its complexity, are drastically affected by the sampling of excitations. We scrutinize them analyzing the various length scales at play in data space. In order to do that, 
let us first focus on excitations where only local changes of configurations are considered (case $(i)$). Given that those only involve the change of a handful of spins, they describe a local modification of the one-dimensional manifold found at $T=0$.
The intrinsic dimension of the corresponding data set has been estimated following a maximum likelihood reasoning in Ref.~\cite{TiagoPRX}, and scales as $I_d\propto \frac{N_r}{\sqrt(N)}$.
The data set does not feature clustering and can be thought of as a (highly anisotropic) multi-dimensional cylinder, of length $\propto \sqrt{N}$ (simply given by the $T=0$ cluster length). We thus refer to the data set as the cylinder cluster.
The width of the cylinder can be computed as follows. Given two configurations  $\vec{\theta}^0$ and $\vec{\theta}^1$, the first representing a fully polarized state with $\theta^0_k =\theta^0$ and the second a state with thermal excitations on top of that polarized state, one gets:
\begin{eqnarray}\label{eq:distance}
d(\vec{\theta}^0, \vec{\theta}^1)  &=& \sqrt{2 \sum_{k-1}^{N} (1 - \cos(\theta^0)\cos(\theta^1_k) - \sin(\theta^0)\sin(\theta^1_k))}   =\nonumber\\
& = & \left\{2 \left[N - \cos(\theta^0)\sum_{k}\cos(\theta^1_k) -\right.\right.\nonumber\\
& - & \left.\left. \sin(\theta^0)\sum_{k}\sin(\theta^1_k) \right] \right\}^{1/2}
\end{eqnarray}
Given that $\vec{\theta}^1$ only differ by $\vec{\theta}^0$ for few spins that are surrounded by spins that are aligned along $\theta^0$, one can rewrite the expression above as:
\begin{equation}\label{eq:d_E}
    d(\vec{\theta}^0, \vec{\theta}^1) = \sqrt{2 \sum_{k} (E^0-E^1_k)}
\end{equation}
where $E_0$ is the average energy of $\vec{\theta}^0$, and 
$E_k$ is the energy associated to the $k$-vertex in $\vec{\theta}^1$. The formula above is providing a direct link between energy of excitations and average distance within the cylinder cluster, that is $d_{\text{cyl}}\simeq{\delta\epsilon}$=  O(1).
}

\mdm{In the absence of topological excitations, the above description will be sufficient to characterize the full data structure. Topological excitations change this situation rather drastically (case $(ii)$). As we show below, analyzing the average distance between configurations, they generate clusters that are geometrically disconnected from the cylinder one, and whose number is given by the dimension of the homology group of the manifold the system is defined on. Such data structure leads to a characterization of the topology-dependence of the intrinsic dimension, which we comment upon at the end of the section.}

\mdm{The key observation is that the average distance $d_{\text{topo}}$ between the cylinder cluster and topological excitations at vanishing energy density is much larger than the estimated $d_{\text{cyl}}$. This is due to the fact that one cannot go from Eq.~\eqref{eq:distance} to  Eq.~\eqref{eq:d_E} as done previously: namely, in the case of topological excitations, each spin configuration varies smoothly with position, so it is not possible to assume that most neighbors have fixed magnetization. Instead, in this case, one gets, from Eq.~\eqref{eq:distance}, that
\begin{equation}\label{eq:d_topo}
\begin{aligned}
    d_{\text{topo}} &\simeq \sqrt{2\sum_k \left[1-\cos(\theta_k)-\sin(\theta_k)\right]} \simeq \sqrt{N}
    \end{aligned}
\end{equation}
due to the fact that the averages of $\cos(\theta_k)$ and $\sin(\theta_k)$ are negligible with respect to the constant term. From a different perspective, this can be understood as a consequence of the lack of long-range order: the latter fixes the average value of $\cos(\theta_k)$ to vanish at finite temperature. While this is of course not strictly true for arbitrary low-energy states, it gives us a rough estimate of the mismatch in scaling in Eq.~\eqref{eq:d_topo}.}

\mdm{The arguments above indicate that topological excitations lead to 'topological' clusters in data space that are distinct from the main cylinder cluster. The number of such additional clusters is also important: in particular, this number scales in general with the number of possible low-energy excitations in the theory (due to the fact that the distance among these clusters is also typically of order $L$). For closed manifolds, such a number is the dimension of the first homology group of the manifold.}

\mdm{The effect of topological clusters on the intrinsic dimension of the manifold can be understood as follows. The intrinsic dimension estimation is still dominated by the cluster cylinder, since the latter contains most of the data points, and at the same time, the nearest-neighbor distance between points in that cluster is smaller than that in the topological ones~\footnote{We note that this picture might be an interesting starting point to investigate the manifold beyond the 2-NN estimator we utilize, e.g., considering fixed-length scale estimators~\cite{macocco2023intrinsic}}. }

\mdm{Summarizing, the predictions of our theory are the followings: $(i)$ complexity is topology independent in the absence of topological excitations; $(ii)$ in presence of topological excitations, complexity decreases proportionally to the dimension of the first homology group of the real-space manifold; $(iii)$ the decrease is associated to clustering, and to a different distribution of nearest-neighbor distances in the sample. }

\mdm{Let us anticipate some concrete expectations based on these predictions. For the case of $g$-torii, one expects the complexity $K(X)$ to decrease proportionally to $2g$. We note that similar scalings also apply to topological phases in quantum statistical mechanics (e.g., two-dimensional lattice gauge theories~\cite{kitaev2006topological}). For the case of closed manifolds, we instead expect a dependence on the number of possible loops - that is, the dimension of the abelianization of the first homotopy group (not its Betti's number). For instance, we expect the same complexity for Klein bottles and 1-torii, despite the fact that their genus is different. }

\mdm{In conclusion, there are aspects of our theory that, at present, we do not know how to address. The first relevant one is, up to which temperatures such a picture holds. The second non-trivial aspect is the role of $N_r$: here, the main challenge is that maximum likelihood arguments (at the basis of the cylinder cluster scaling) are known to be not always accurate in terms of capturing the role played by the data set size. To clarify these points, and to corroborate our full theory picture, we rely on numerical experiments. }

\subsection{Numerical experiments}\label{sec:Numerics}

We now provide numerical simulations to support our theory. We consider a two-dimensional square lattice whose boundary conditions are controlled such that topologically different manifolds are realized. In Fig.~\ref{fig:fig2}a) we illustrate how to build a 2-torus \vv{($T^2$)} with appropriate constraints on the boundaries of two square lattices. Equivalently, one could start from an octagon and apply periodic boundary conditions on opposite sides. The extension to generic $g$-torii is straightforward starting from regular polygons with $4g$ sides or using $g$ square lattices. In practice, we start from a square lattice and divide it in vertical stripes instead of considering $g$ square lattices. In the Supplemental Material Sec.~\ref{sec:SvS} we show numerical evidence of the equivalence of the two approaches.
Non-orientable manifolds like the Klein bottle, the Moebius strip, and the real projective plane (RPP) can be realized with the appropriate boundary conditions on a single square lattice (see Fig.~\ref{fig:fig3}a). 

In order to sample the configurations according to the partition function we use Wolff's algorithm \cite{Wolff1989}. The data sets are defined as a collection of $N_r$ points, such as in Eq.~\eqref{eq:datasets}, each corresponding to a given \vv{configuration} of spins as specified in the previous section. We take particular care of generating decorrelated configurations, in order to avoid any possible bias of the sampling. To do so we choose a suitable interval between configurations that are saved during the Monte Carlo steps. We note that, to the best of our knowledge, for some of these manifolds, numerical simulations were not reported before, so we verified that autocorrelations are not significantly changed by the manifold topology.

\subsubsection{Orientable closed manifolds}

The first step of our analysis is the comparison of different $g$-torii. The number of non-trivial loops of a genus $g$ torus (T$^{g}$) is $2g$, since its first-homotopy group is $\pi_1(\mathrm{T}^{g}) = \bigoplus_{1=1}^{2g} \mathbb{Z}$.
We expect, in a phase with no topological features, that the complexity should not show any dependence on $g$.
Conversely, in the BKT phase, we expect it to be dependent on the particular manifold taken into account.
Our theory predicts that the complexity $K(X,g)$ of the data set $X$ as a function of the genus $g$ decreases with $g$ monotonically, up to finite volume effects. In order to cope with the latter, we utilize a phenomenological finite volume ansatz as:
\begin{equation}\label{eq:genus_scaling}
    K(X,g) = K_0 - a_0 \cdot g \mathrm{e}^{-a_1 \sqrt{g}}
\end{equation}
which smooths out contributions at large $g$, since those are more sensitive to finite volume effects scaling roughly as $V\simeq g^2$~\footnote{It might be possible to refine this ansatz incorporating eventual anisotropies in the geometry of the system. However, for the system sizes we are interested in, it would be challenging to make concrete claims (due to the challenging scaling of our $I_d$ estimators versus the embedding dimension), on that, so we leave this aspect to future investigations.}.
On the right-hand side, we include a topologically independent term $K_0$ and a topologically dependent term, that will capture the TKC.

In Fig.~\ref{fig:fig2} we show the estimated value of the intrinsic dimension $I_d$ as a function of temperature for different values of the genus $g$ (b) and as a function of the genus $g$ for different values of the temperature (c).
The system taken into account is a square lattice with linear dimension $L=72$ and the number of configurations \vv{of} the data set is $N_r=5000$.  We observe that the curve are monotonically increasing with temperature as expected. \vv{The presence of the phase transition is not detected by the intrinsic dimension, differently to Ref.~\cite{TiagoPRX}, because the number of configurations considered is lower; however, this fact is irrelevant to our analysis as we show below.}

The most striking signal in Fig.~\ref{fig:fig2}b) is that, below $T/J\lesssim 1.2 /k_B$, the \vv{intrinsic} dimension decreases as a function of the genus, and the distance among the curve remains constant, at least deep in the topological phase, where our reasoning applies. We scrutinize such genus dependence in Fig.~\ref{fig:fig2}c). The agreement between the points and the ansatz (curves obtained via a best fit based on Eq.~\eqref{eq:genus_scaling}) is excellent for low temperatures. Oppositely, it worsens with increasing $T$ and cannot describe $I_d$ above the phase transition where the topological properties are not relevant any longer.
We argue that in the thermodynamic limit, the curves for different values of the genus should overlap for all temperatures above the critical one $T>T_{BKT}$. To check this assumption we calculate $I_d$ for $T=1.4>T_{BKT}$ and observe that the values of the different curves are compatible within the error bars, as shown in Fig.~\ref{fig:fig2}b).

\subsubsection{Non-orientable and open manifolds} 

While for orientable manifolds we gave our predictions based on the number of independent non-trivial loops, in the case of non-orientable ones we need to properly extend our diagnostic. This can be done by looking at the first homology group of the manifold, i.e. the abelianization of their first homotopy group.
We expect that: \emph{i)} data sets defined on spaces with the same dimension of the homology group will be characterized by the same Kolmogorov complexity irrespectively of their genus and their orientability; \emph{ii)} the complexity may be strongly affected by the number of open boundaries in finite-size settings; \emph{iii)} in the phase with no topological properties the complexity shall show no scaling with the topological features of the manifolds.

\begin{figure*}
    \centering
    \includegraphics[width=\linewidth]{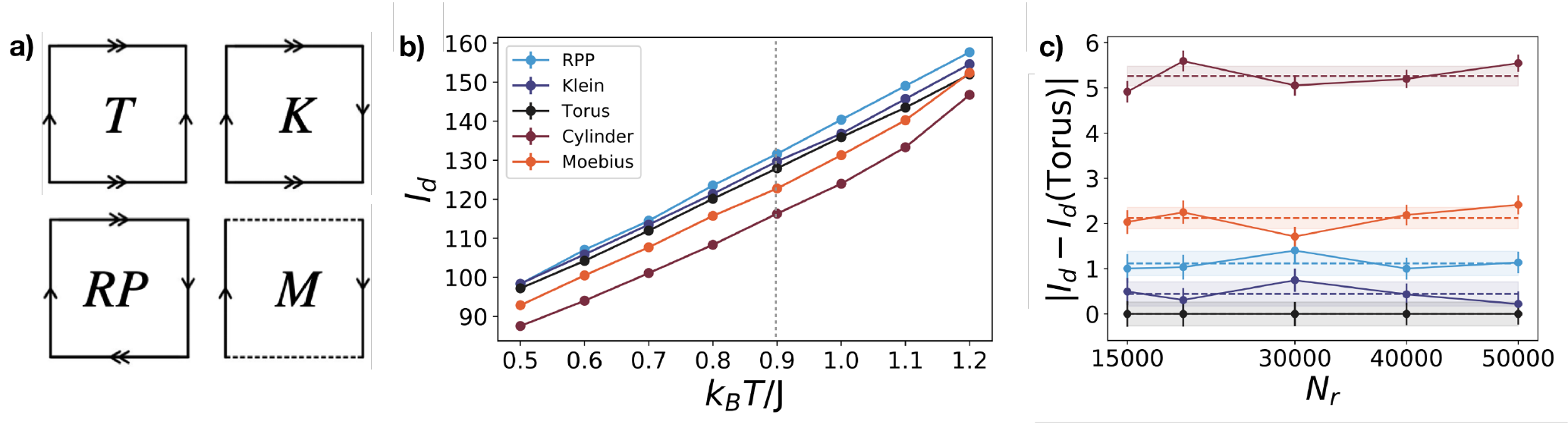}
    \caption{{\bf Comparison between RPP, Klein, torus, cylinder and M\"obius topologies}. We consider the XY model on a square lattice. a) Diagram of the construction of torus (T), Klein bottle ($K$), real projective plane ($RP$) and M\"obius strip ($M$). \vv{Edges} with the same arrows must be connected with periodic boundary conditions. \vv{Edges} with arrows in opposite directions denote anti-periodic boundary conditions. Dashed lines stand for open boundary conditions. b) results of $I_d$ as a function of temperature across the BKT phase transition, on a square lattice with linear dimension $L$, from top to bottom we have results for RPP, Klein $\sim$ Torus, Möbius strip, Cylinder. We average our results over 10 simulations with $N_r=5000$. \vv{The error bars are estimated as the standard deviation of the mean.} \vvv{ The dashed vertical line marks the critical point.} c) $|I_d-I_d(\mathrm{Torus})|$ at $K_bT/J=0.1$ as function of $N_r$ for a square lattice with linear dimension $L=48$. The dashed lines correspond to the average of $I_d$ over $N_r$ for the given manifold. Their error bars are represented by the \vv{shaded regions} of the same \vv{colors}. From top to bottom we have results for Cylinder, Möbius strip, RPP, Klein $\sim$ Torus. }
    \label{fig:fig3}
\end{figure*}

\paragraph*{Homology and complexity. -}
In Fig.~\ref{fig:fig3}b) we plot $I_d$ of the XY model, on a square lattice with linear dimension $L=72$, calculated from $N_r=5000$ configurations of spins as in Eq.~\eqref{eq:datasets}. We consider the cases of spins arranged on a real-projective plane \vv{(RPP)}, cylinder, torus, Klein bottle, and M\"obius strip (schematically depicted in Fig.~\ref{fig:fig3}a)). \vv{The corresponding homology groups are written in Tab.~\ref{table:1}.}
\vvv{\begin{table}
\centering
\begin{tabular}{|c|c|c|c|c|}
    \hline
    \multicolumn{3}{|c|}{Closed manifolds} & \multicolumn{2}{|c|}{Open manifolds} \\
    \hline
     RPP &  Torus & Klein bottle& Cylinder & M\"obius strip \\
\hline
$\mathds{Z}$ &  $\mathds{Z}\oplus \mathds{Z}$ & $\mathds{Z}\oplus \mathds{Z}_2$ & $\mathds{Z}$ & $\mathds{Z}$\\
\hline
    \end{tabular}
\caption{Homology groups of the topologies considered}
\label{table:1}
\end{table}}
This choice of topologies allows us to consider the influence of homotopy \vv{groups} and boundaries separately.
According to our definition of TKC, we utilize \vv{the 1-torus} as a reference: the cylinder allows us to investigate the role of boundaries in the estimation of $I_d$, while the others are non-orientable, closed manifolds which are the main focus of the present section.

From Fig.~\ref{fig:fig3}b), few qualitative features are appreciable. First, the presence of boundaries in the manifold decreases the value of $I_d$ consistently. \vv{This is clear from the result of the M\"obius strip (one boundary) and cylinder (two boundaries), that have the same homology group}. This can be explained as due to the constraint enforced by the presence of an edge that amounts to adding more labels ($W_j$) to the data. \vv{Secondly we see that torus and Klein bottle have compatible $I_d$ for the whole temperature range investigated, in agreement with our reasoning in the previous section (Sec.~\ref{sec:windingnumberandinformation}).}
In order to quantify such connection, in Fig.~\ref{fig:fig3}c), we plot the absolute value of the topological Kolmogorov complexity as a function of $N_r$ for a square lattice with $L=48$ at low temperature $k_bT/J=0.1$. For each topology, we also estimate an average value of the TKC across our range of $N_r$: this is indicated by dashed lines, and the corresponding estimated error is indicated by a \vv{shaded} area. 

Two main observations are in order. The first is about the relative behavior of the closed manifolds: there is a sharp difference between RPP and 1-torus. In particular, their TKC difference is compatible with 1 regardless of the sampling. \vv{This is remarkable as, e.g. for $N_r=50000$, the overall intrinsic dimension exceeds 100}. Oppositely, the difference between Klein bottle and 1-torus is always small and \vv{it is compatible with 0, for 3 values of the sampling}. \vv{Thus we can state that} the TKC of a partition function supporting topological excitations depends on the homology of the manifold (and, e.g., not only on its genus).
The second observation is that, as expected, the presence of edges is strongly affecting the TKC. This is evident from the fact that \vv{the M\"obius strip, cylinder and RPP have very distinct TKC, despite their homology groups having the same dimension or being exactly the same}.

\begin{figure}
    \centering
    \includegraphics[width=\linewidth]{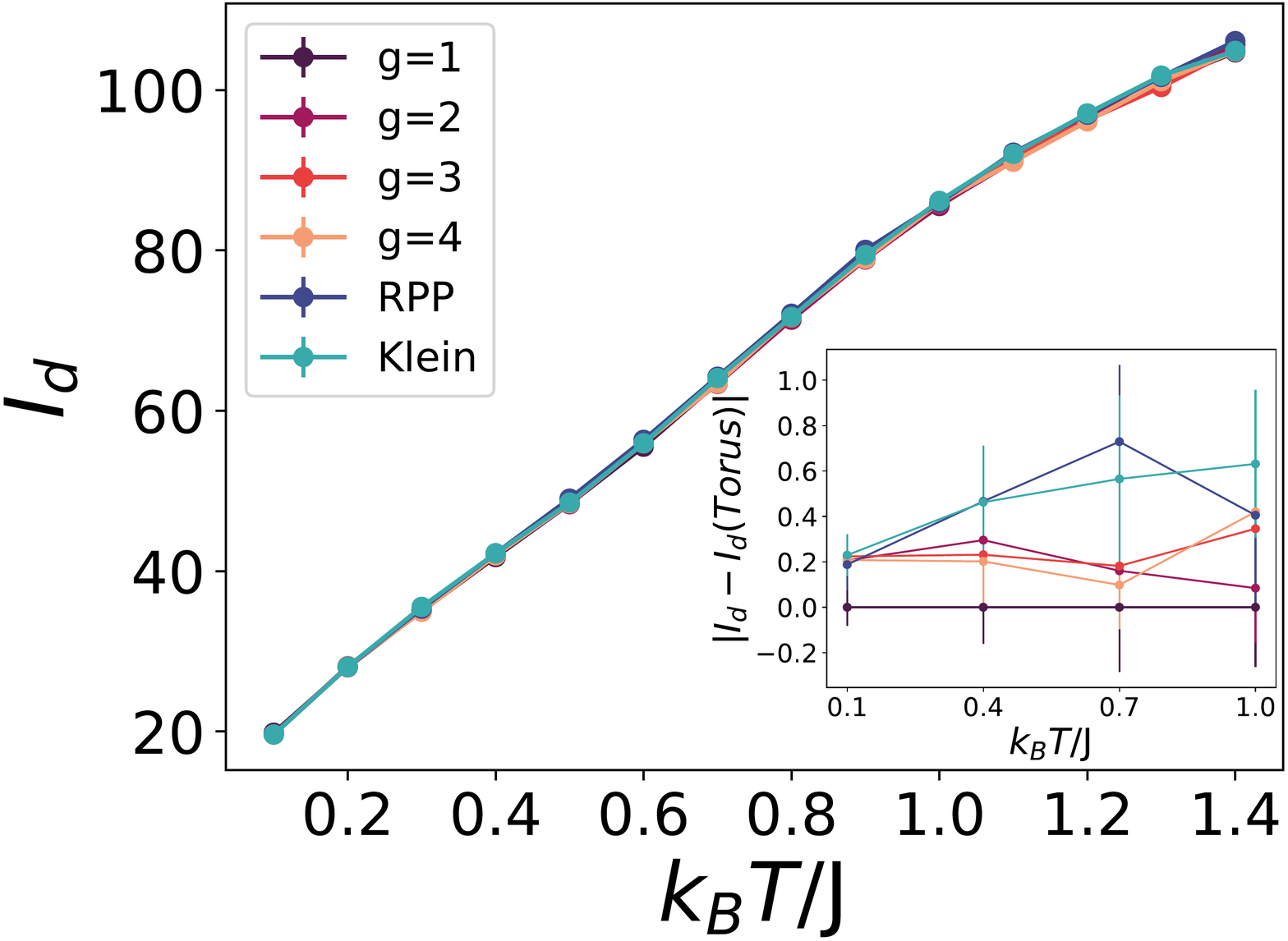}
    \caption{\vv{{\bf Plot of $I_d$ for the Heisenberg model}. We consider a square lattice with linear dimension $L=72$. We consider $N_r=5000$ and $I_d$ is estimated as average over $10$ simulations. The error bars are estimated as the standard deviation of the mean.  It is evident the absence of any topological features. In the inset, we show the difference of $I_d$ of all the topologies with respect to the reference case of the torus. We observe the values are compatible within 1 standard deviations of the mean.}}
    \label{fig:Heisenberg}
\end{figure}

\vv{ We now contrast the XY results with what we observe for the case of the Heisenberg model, depicted in Fig.~\ref{fig:Heisenberg}. In that case, it is known that no transition is present as a function of temperature. We observe that the values of $I_d$ are compatible for any topology. We attribute this to the fact that the model is topologically trivial and shows a clearly different behavior with respect to the XY model. In the inset of Fig.~\ref{fig:Heisenberg} we plot $|I_d-I_d(\mathrm{Torus})|$ for a few values of temperature to highlight that the curves are compatible within error bars. In the Supplementary Material (Sec.~\ref{sec:Ising}) we also show the results for the Ising model. The conclusions we can draw are similar. There is not a clear trend in the value of the intrinsic dimension with respect to topology. Indeed, all the values -- for different topologies -- are compatible within error bars since both phases of the Ising model are topologically trivial.}

\subsubsection{Conditional connectivity: a local order parameter in data space}\label{sec:CC}
\begin{figure}
    \centering
    \includegraphics[width=0.95\linewidth]{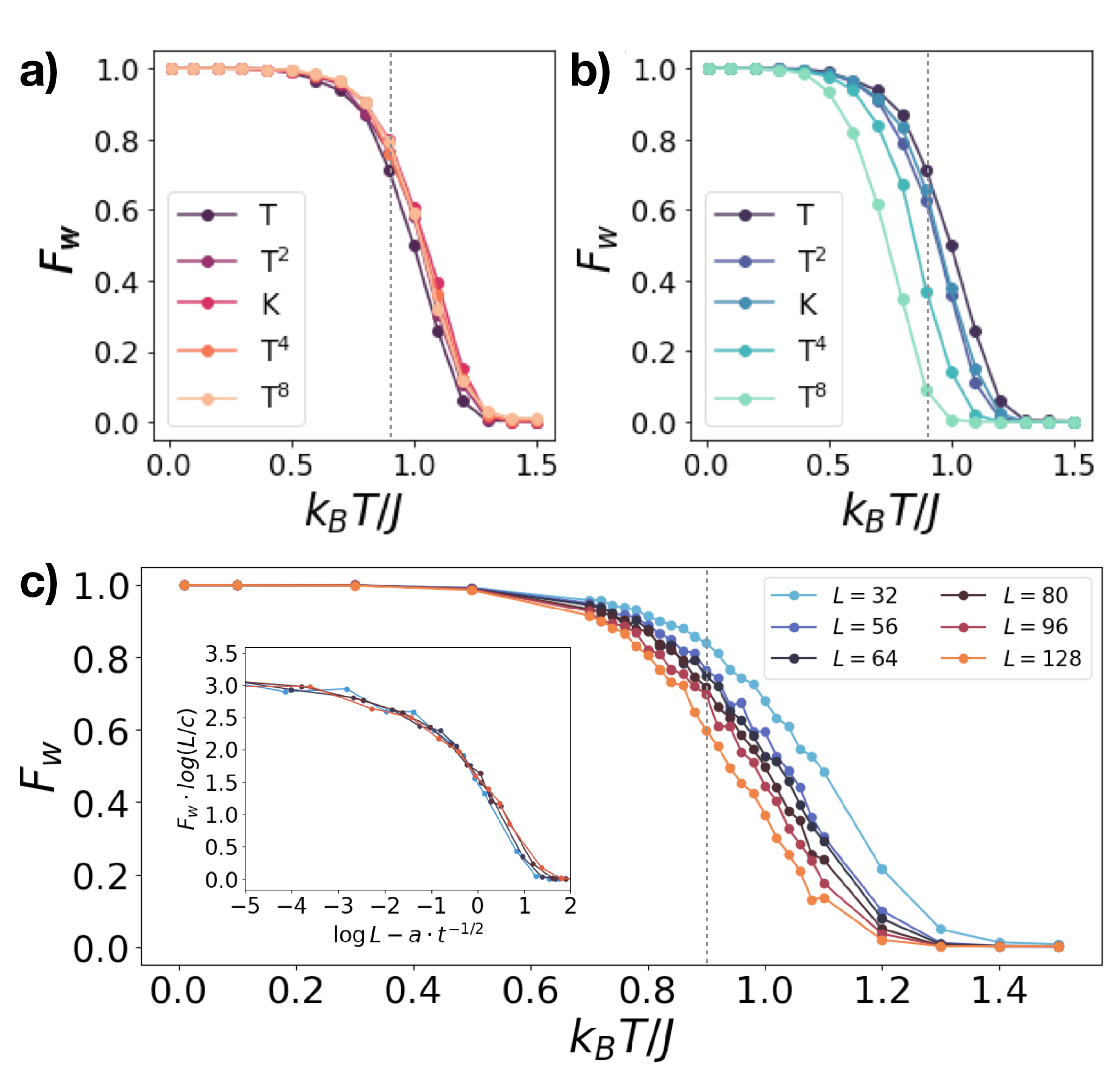}
    \caption{{\bf Fraction of points in data space whose first two nearest neighbors have the same winding number $F_{\mathbf{w}}$}. In panel a) we consider only two non-trivial loops. In panel b) we consider every possible non-trivial loop on the manifold. The crossover point moves (from right to left) depending on the topology according to the legend. The square lattice has linear dimension $L=72$.
    In panel c) we consider $F_{\mathbf{w}}$ for the 1-torus at different system sizes $L^2$, going from $L=32$ (top curve) to $L=128$ (bottom curve).  These results indicate how, below the critical temperature \vvv{(labeled by the dashed vertical lines)}, the {\it local} (connectivity) properties of the data set become severely constrained by {\it global} properties (winding numbers) of the configurations sampled. This is a local order parameter in data space that has a universal behavior at the transition, as shown in the inset. Here $t=K_b(T-T_c)/J$, with $T_c$ value set at $K_bT_c/J=0.8933$, and $a=1.767 \pm 0.003$ and $c=0.721 \pm 0.003$, estimated through a data collapse. }
    \label{fig:windings}
\end{figure} 

In order to validate the argument above, we need to identify testable predictions that connect topological excitations in real space with the data structure, and that can be independently verified when investigating partition function samples. Informed by our analysis above, we focus on the role of winding numbers in the data structure. 

The winding number $w_
{\Gamma}$ is a key quantity for studying topological systems. 
In a spin system, it is an estimate of the number of times the spins wind following a given path $\Gamma$ along the manifold.
In the case of the XY model, where the $i$-th spin is represented by a single angle $\theta_i$, it can be written as follows:
\begin{equation}
    w_{\Gamma}=\frac{1}{2 \pi} \sum_{i=1}^{L_{\Gamma}}\Delta \theta_{i,i+1},
\end{equation}
where $\Gamma$ is comprised of $L_{\Gamma}$ sites, and $\Delta \theta_{i,i+1}=\theta_{i+1}-\theta_i$ is rescaled into the range $(-\pi,\pi]$.
As shown in Fig.\ref{fig:summary}(iv) the possible independent paths on which winding numbers can be calculated depend on the manifolds themselves.
We define as $\mathbf{w}=\{ w_{\Gamma_1},\dots,w_{\Gamma_k} \}$ the collection of $w_{\Gamma}$'s calculated on the paths $\Gamma_1,\dots,\Gamma_k$.
For instance, in the case of a torus ($T$) and a 2-torus ($T^2$), one has $\mathbf{w}(T)=\{w_{a},w_{b}\}$ and $\mathbf{w}(T^2)=\{w_{a_1},w_{b_1},w_{a_2},w_{b_2}\}$, respectively (see Fig.\ref{fig:summary}(iv)).
Remarkably, $\mathbf{w}$ is a key metric for understanding the clustering structure of configurations in phase space because winding numbers capture the topological nature of excitations. Namely, above the BKT transition, vortex-antivortex pairs are unbounded, resulting in a wide range of different $\mathbf{w}$ values among sampled configurations. However, in the quasi-long-range-order regime (below $T_{BKT}$), the majority of configurations have the same $\mathbf{w}$, i.e. for a torus topology $\mathbf{w}= (0,0)$. 

According to the rationale above, the physical properties of the BKT phase transition strongly affect the connectivity between neighboring configurations.
In particular, the fraction of points whose first and second neighbors have the same $\mathbf{w}$, that we dub \emph{conditional connectivity} $F_{\mathbf{w}}=N_{\mathbf{w}}/N_r$, must be informative of the transition.
Here $N_r$ is the total number of configurations sampled and $N_{\bf{w}}$ is the number of configurations whose first two neighbors have the same $\mathbf{w}$.
Evidently, $F_{\mathbf{w}}$ is close to zero above the BKT transition (where configurations exhibit a wide range of $\mathbf{w}$), but is equal to 1 in the topological phase (where the values of $\mathbf{w}$ are constrained due to the topological nature of the excitations and the configurations appear as clusters in data space).

In Fig.~\ref{fig:windings}a) we show $F_{\mathbf{w}}$ in the case where only two paths on the lattice are taken into account for each manifold, that is to be compared with Fig.~\ref{fig:windings}b) where all the possible non-trivial paths are considered.
We observe that the two figures display the same qualitative behavior, showing a transition near the critical temperature, but they are quantitatively different. This is expected since the larger the number of possible independent windings the more constrained is the system in the BKT phase.
In Fig.~\ref{fig:windings}c) we show $F_{\mathbf{w}}$ for the case discussed in a) but for different system sizes.  In the inset we perform a collapse of the curve with estimated parameters $c=0.72(1)$ and $a=1.76(7)$. To do so we use a suitable mesh for the parameters, fixing the critical temperature $k_BT_c/J=0.8933$ \cite{sandvik2010computational}. Then we calculate the scaling variables 
\begin{equation}
x=\ln L -a t^{-1/2},\;\; y=F_{\mathbf{w}} \ln(L/c)
\end{equation}
and choose a parametric function hypothesis $f(x,a,c)$ to be compared with $y$. Here $t=(T-T_c)/T_c$.
We use the Levenberg-Marquardt algorithm to compute the best fit for each choice of $a,c$, and $f$, where $f$ is considered to be a polynomial with varying degree $k=5,6,7,8$. Our estimates for $a$ and $c$ are calculated as the average of all the best fits and their errors are the standard deviation of the latters.
We observe that the conditional connectivity $F_{\mathbf{w}}$ displays a universal behavior. It can be considered as an order parameter in the data space and, as argued before, is closely related to the fact that the Kolmogorov complexity displays a topological contribution in the BKT phase. In fact, the local connectivity (as estimated by $I_d$) is strongly constrained by the global properties of the manifold, which are expressed by the behavior of $F_{\mathbf{w}}$.
This discussion allowed us to make two clear statements that we want to summarise here:
i) The connectivity between neighboring configurations is strongly related to the physical properties of the BKT transition. Remarkably, nearest-neighbor configurations have identical physical properties (winding numbers) when the system is in the ordered (topological) phase. Conversely, in the disordered phase, the first and second neighbors physical properties are entirely
random as witnessed by $F_{\mathbf{w}}$ approaching zero; ii) changing the number of $w$'s calculated and compared, the transition point ($F_{\mathbf{w}}$ going to 1) moves as more constraints are applied on the configurations. Hence it should be expected to have different behaviors for the Kolmogorov complexity $K(X)$ on different manifolds, as it has clearly shown before. 

\section{Discussion and conclusions}

We have proposed a theory that connects topological properties of classical statistical mechanics, to the Kolmogorov complexity of the corresponding statistical sampling of the many-body state according to the problem partition function. The original approach we proposed is inspired by topological field theory: take a manifold, sample the partition function on the model defined on the manifold itself, and get an invariant. By gauging such a procedure on 1-torus, we have defined a topological Kolmogorov complexity associated to a given problem, on a given manifold. Based on the connection we establish between topological properties in real space, and data structures in configuration space, the overall conclusion of our theory is that the more constrained a many-body system is by topological effects, the lower the Kolmogorov complexity of its sampling will be.

We have then addressed our theory in the framework of classical statistical mechanics models - Ising\vv{, Heisenberg} and XY defined on square lattices. In the Berezinskii-Kosterlitz-Thouless phase of the XY model, the complexity of the partition function strongly depends on the homology of the target manifold, demonstrating a strong, qualitative connection between topology and complexity. Most importantly, the topological Kolmogorov complexity attains universal (in the sense of sampling-independent) values in the BKT phase, that are suggestive of its direct link to the winding of spins in the manifold. While some of these values appear compatible with integer TKC, we believe a more extended study is required here, that shall address larger datasets from the ones we consider here. This will require the development of new methods for intrinsic dimension estimation, in particular, with respect to errors. In our numerical experiments, we have also observed a strong dependence on complexity with respect to boundaries. Finally, at high temperatures, and in the Ising \vv{and Heisenberg models}, we observe no connection between topology and complexity, as expected. 

Within the XY model, our theory allows us to track the origin of the relation between complexity and topology by analyzing the interplay between metric distances and winding numbers. In particular, we identify an order parameter - conditional connectivity - that, while fully local in data space, correctly predicts the topological transition taking place in real space. The order parameter satisfied phenomenological scaling relations, a further proof of its relation with the BKT transition. This \vv{interpretation} might be useful in understanding how learning methods such as dynamical maps and persistent homology have found very convincing applications on the XY model \cite{Scheurer2019,Tran2021,Lucini2022,otsuka2023comprehensive}, by specifically exploiting information about winding numbers. In fact, these methods might have indirectly detected the simple order parameters we proposed. It is worth commenting on the fact that, while exotic topologies might seem hard to realize in general, they are an experimental reality in quantum simulators. In fact, those can be achieved in a variety of manners, including synthetic dimensions~\cite{boada2012quantum,boada2015quantum}, atom configurations, and atom rearrangements in optical tweezers, the last one being already experimentally demonstrated~\cite{bluvstein2022quantum}.

Summing up, our theory and numerical experiments support the fact that real-space topological information becomes 'local' in data space. This last conceptual insight can serve as a starting point for the exploration of the interplay of complexity and topology beyond equilibrium classical statistical mechanics. One clear avenue is off-equilibrium physics, where the same toolbox we used here shall be in principle directly applicable.
Another horizon are quantum equilibrium systems. In that context, the connection between information and topology has found profound application in topological matter - like, e.g., the formulation of topological entanglement entropy~\cite{kitaev2006topological}. However, such quantities are typically not amenable to experimental test at large scale, as they require the knowledge of the full system wave function. A statistical approach like the one pursued here could reveal signatures of topological order and/or topological effects even in the presence of limited, but statistically significant, sampling, as long as experiments are able to dynamically tune topology (as demonstrated, e.g., in Ref.~\cite{nogrette2014single,ebadi2022quantum}). It is however important to stress that, while our work provides basic tools in this direction (for instance, the partition functions we study can be thought of as path integrals of one-dimensional fluids with fractionalized excitations), attacking the quantum regime necessarily requires a more convoluted approach. This is not only due to the fact that topology has different incarnations in that setting (e.g., topological order, band structures, and critical phenomena) but, most importantly, to the fact that quantum systems provide data sets in multiple basis. It is an open question to determine how a clever combination of those, possibly together with randomized sampling \cite{elben2022randomized}, would lead to a direct connection between topology and complexity.

\section*{Aknowledgements}
We acknowledge useful discussions with R. Panda, R. Verdel and E. Vicari, and thank A. Angelone, M. Heyl, S. Pedrielli and M. Schmitt for collaborations on related topics. V.V. acknowledges discussions with P. Pegolo and G. Giachetti. The work of M.~D. and V. V. was partly supported by the ERC under grant number 758329 (AGEnTh), by the MIUR Programme FARE (MEPH), by the PNRR MUR project PE0000023-NQSTI.
V.V. acknowledges computing resources at Cineca Supercomputing Centre through the Italian SuperComputing Resource Allocation via the ISCRA grants ICTP21\_CMSP and ICTP22\_CMSP. M.D. was partly supported by the Munich Institute for Astro-, Particle and BioPhysics (MIAPbP) which is funded by the Deutsche Forschungsgemeinschaft (DFG, German Research Foundation) under Germany´s Excellence Strategy – EXC-2094 – 390783311.
The work of TMS was supported by the European Research Council (ERC) under the European Union’s Horizon 2020 research and innovation programme (grant agreement No. 853443).

\appendix
\section{Ising model results}\label{sec:Ising}
\begin{figure}
    \centering
    \includegraphics[width=0.9\linewidth]{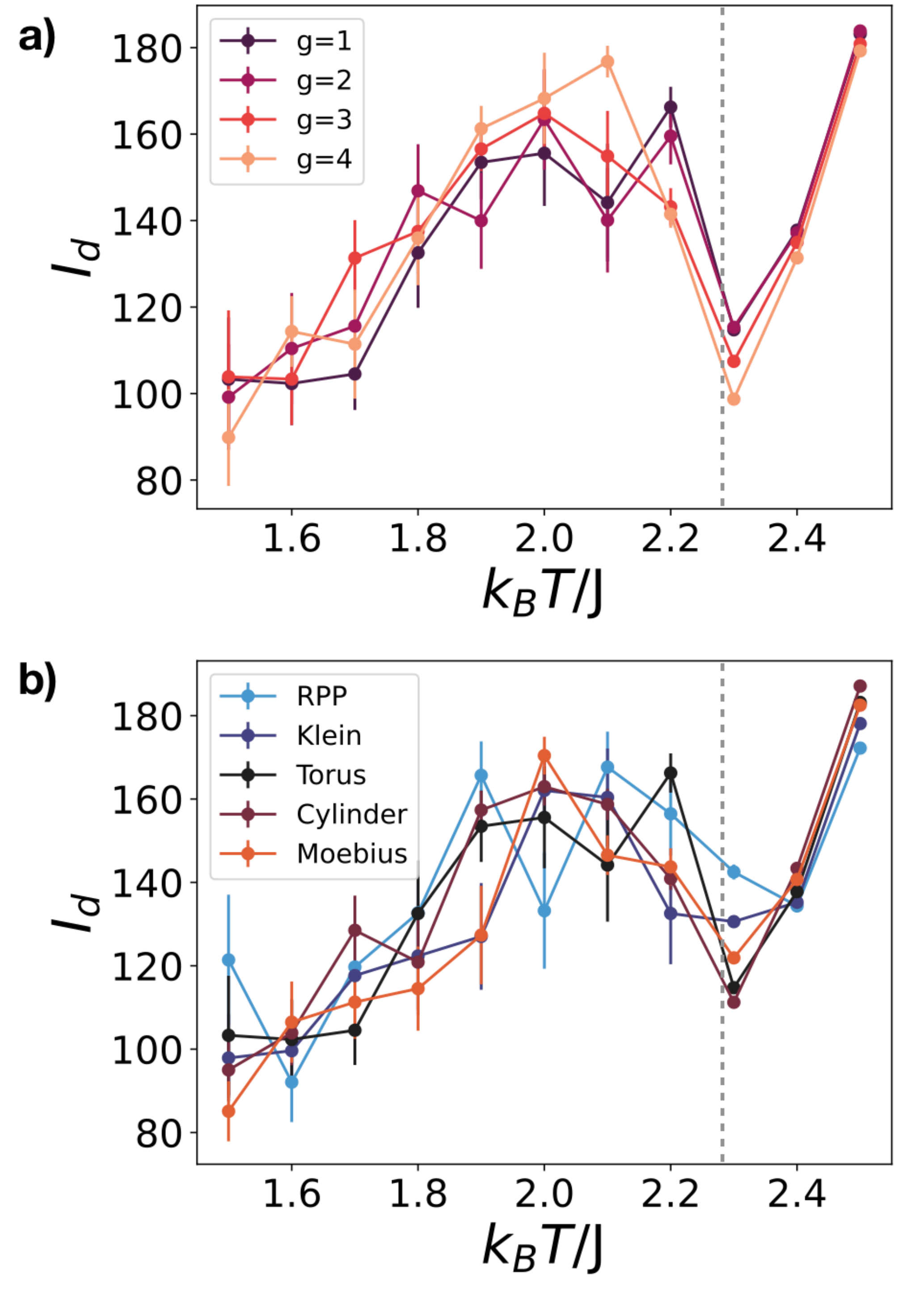}
    \caption{{\bf Plot of $I_d$ for the Ising model}. We consider a square lattice with linear dimension $L=72$. We consider $N_r=10^4$ and $I_d$ is estimated as average over $10$ simulations. \vv{Value of $I_d$ for different topologies. Both for a) and b) We observe all the values are compatible within the error bars as both phases of the Ising model are topologically trivial.} The critical point is denoted with a dashed vertical line.}
    \label{fig:Ising}
\end{figure}

\vv{We here contrast the XY results with what we observe for the case of the Ising model. In particular, we show  $I_d$ as a function of the temperature for Torii of different genus $g$ (Fig.~\ref{fig:Ising}a)), and other topologies (Fig.~\ref{fig:Ising}b)). We consider the Ising model on a square lattice with L=$72$.  We
take $N_r = 10^4$ snapshots and $I_d$ is estimated as an average over 10 simulations.
We observe that $I_d$ clearly shows a local minimum near the critical temperature as expected but it does not show dependence on the topology of the manifold the systems lie on. We observe all the values are compatible within the error bars as both phases of the Ising model are topologically trivial. Moreover, finite size effects play here a strong role at the transition point, where most configurations differ in a way that is distinct from the XY case. Here, the complexity of Klein and torus -- that are within error bars for the XY case given their homotopy groups -- widely differ.}

\section{Pareto Distribution check} 
\begin{figure}
    \centering
    \includegraphics[width=0.7\linewidth]{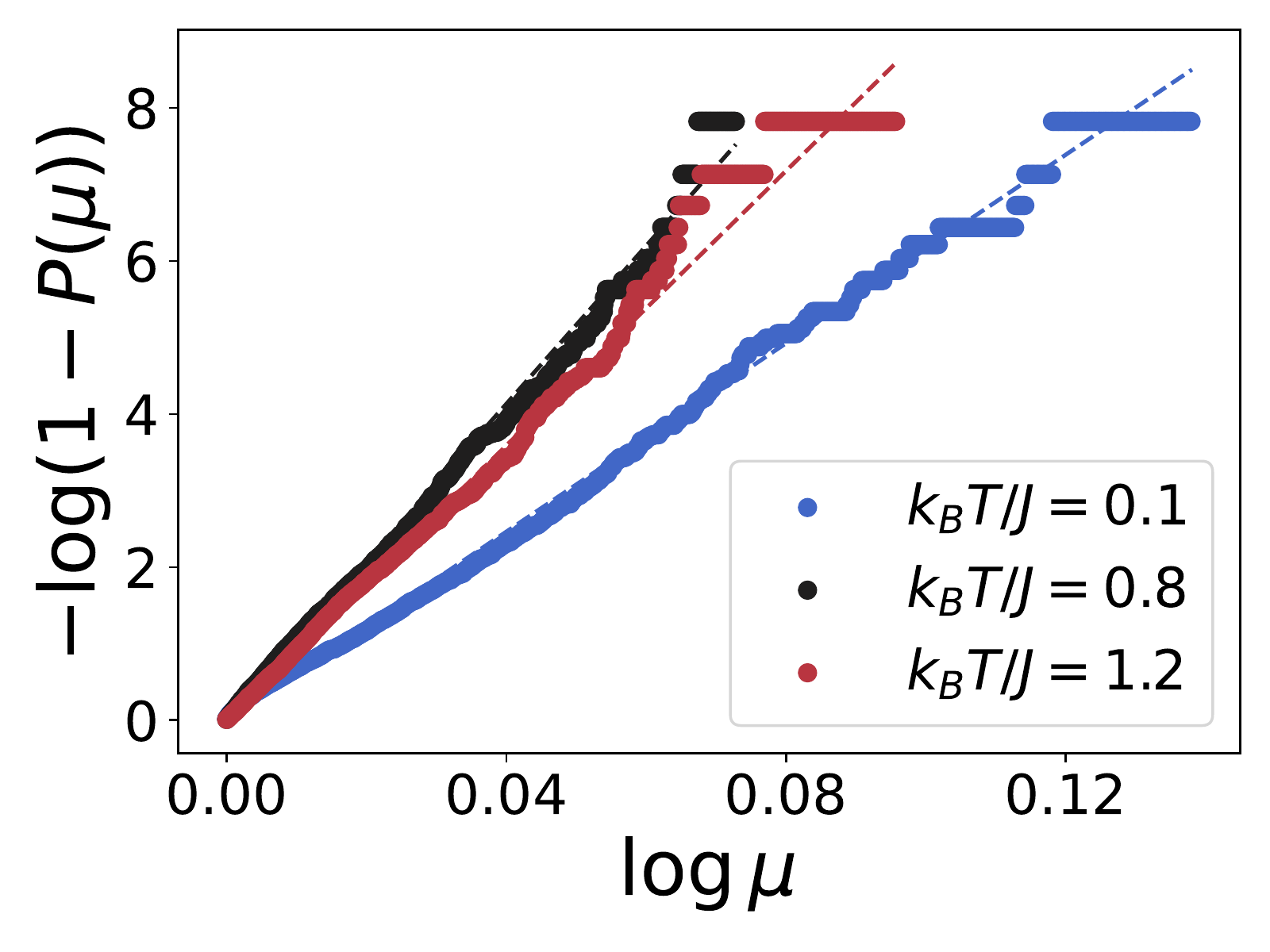}
    \caption{{\bf Check of the Pareto distribution}. Here we consider a square lattice with $L=72$ and $N_r=2500$. Here $t=k_b T/J$.}
    \label{fig:Paretocheck}
\end{figure}

Starting assumption of the TWO-NN method is that the ratio $\mu_i=r_i^{(2)}/r_i^{(1)}$, of the next-nearest- and nearest-neighbor is Pareto distributed, namely that $f(\mu)=I_d \mu^{-I_d-1}$. 
Here we check this assumption for a few cases of the XY model, we fix temperature $k_bT/J=0.1$, $k_bT/J=0.8$ and $k_bT/J=1.2$ and compute $I_d$. Then we compare the cumulative distribution corresponding to the estimated $I_d$ and the one computed from the distribution of distances in the data set.
We obtain the plots in Fig.~\ref{fig:Paretocheck}. It is evident a perfect agreement between the curve (exact cumulative distribution function) and the points (cumulative distribution obtained from the samples configurations).\\

\section{Strings vs Squares}\label{sec:SvS}
\begin{figure}
    \centering
    \includegraphics[width=0.8\linewidth]{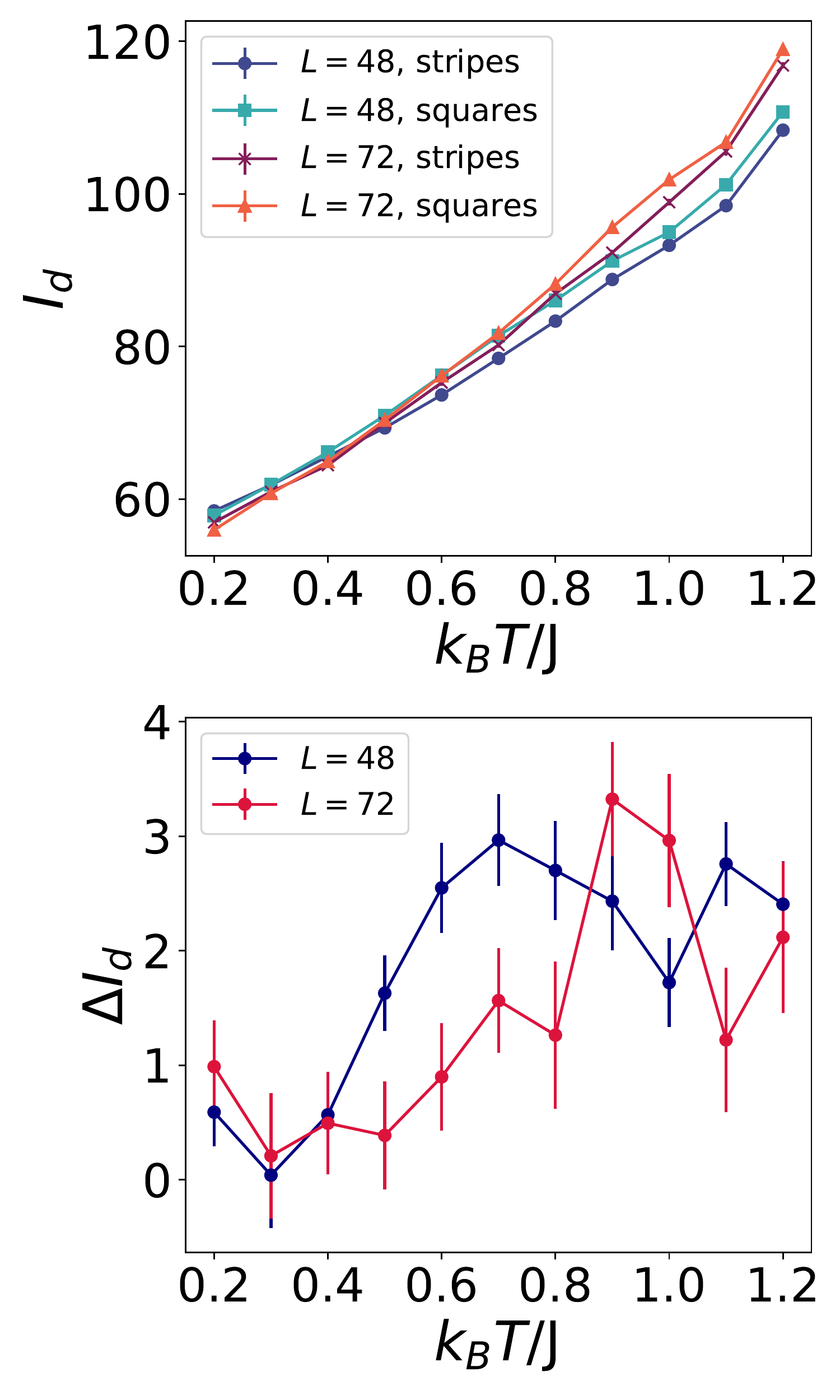}
    \caption{{\bf Comparison between the different procedures for getting a $4$-torus}. Comparison between 4 squares lattices with $L=24$ and rectangular lattices with linear dimensions $48$ and $12$; and between square lattices with $L=32$ and rectangular lattices with linear dimensions $72$ and $18$.}
    \label{fig:squaresvsstripes}
\end{figure}

In the main text, we have described the method we use for computing $I_d$. In a nutshell, we sample configurations via Wolff's algorithm and realize the different manifolds appropriately tuning the boundary conditions.\\
We observe that in the case of the $g$-torii the easiest way to do so, from a computational point of view, is to consider a single square lattice, to divide it in $g$ stripes, and to apply the necessary periodic boundary conditions, instead of considering $g$ square lattices. 
This allows us to consider easily the same number of lattice sites, i.e. the same lattice volume, for each genus $g$ and to realize also odd-$g$ torii.
In this section, we present the comparison, for $g=4$, between $I_d$ calculated using 4 square lattices or a single square lattice divided into 4 stripes.
In the upper panel of Fig.~\ref{fig:squaresvsstripes} we present the comparison between $48^2$ sites and $72^2$ sites for the two different geometries, for the XY model.
We observe that even though they differ around the critical temperature (which is shifted towards $k_bT/J=1$ due to finite size corrections), increasing the system size entails that the agreement between the two geometries grows.
This is due to the fact that the correlation length grows at the transition point such that finite size effect are much more relevant. However, we observe that these effects at the transition do not matter in view of the purpose of the main text since our understanding applies at the topological phase.
In the lower panel of Fig.~\ref{fig:squaresvsstripes} we show explicitly the difference in $I_d$ of the two approaches $\Delta I_d$. We observe $\Delta I_d$ starting growing later in the case of larger system sizes, testifying that  $\Delta I_d>0$ is a finite size effect.

\bibliography{biblio}

\end{document}